\documentstyle[aps,12pt]{revtex}

\widetext
\draft
\tighten
\begin{document}

\preprint{EFUAZ FT-96-38-REV}

\title{Gauge Transformations For Self/Anti-Self Charge Conjugate
States\thanks{Submitted to ``Acta Physica Polonica B".}}

\author{{\bf Valeri V. Dvoeglazov}}

\address{Escuela de F\'{\i}sica, Universidad Aut\'onoma de Zacatecas\\
Apartado Postal C-580, Zacatecas 98068 Zac., M\'exico\\
Internet address: VALERI@CANTERA.REDUAZ.MX\\
URL: http://cantera.reduaz.mx/\~~valeri/valeri.htm}

\date{First version: December 1996. Revised version: June 30, 1997.}

\maketitle

\begin{abstract}
Gauge transformations of type-II spinors are considered in the
Majorana-Ahluwalia construct  for self/anti-self charge conjugate
states.  Some speculations on the relations of this model with
the earlier ones are given.
\end{abstract}



Recently, new constructs in the $(1/2,0)\oplus (0,1/2)$ representation of
the Lorentz group have been proposed~\cite{DVA,Ziino,Robson,DVO1,DVO2}.
The one of their surprising features is the fact that dynamical equations
in these formalisms take eight-component form. As shown in
refs.~\cite{DVA,DVO1,DVO3} the Majorana-McLennan-Case construct for
self/anti-self charge conjugate states leads to the equations:
\begin{mathletters}
\begin{eqnarray}
i \gamma^\mu \partial_\mu \lambda^S (x) - m \rho^A (x) &=& 0 \quad,
\label{11}\\
i \gamma^\mu \partial_\mu \rho^A (x) - m \lambda^S (x) &=& 0 \quad;
\label{12}
\end{eqnarray}
\end{mathletters}
and
\begin{mathletters}
\begin{eqnarray}
i \gamma^\mu \partial_\mu \lambda^A (x) + m \rho^S (x) &=& 0\quad,\\
\label{13}
i \gamma^\mu \partial_\mu \rho^S (x) + m \lambda^A (x) &=& 0\quad.
\label{14}
\end{eqnarray}
\end{mathletters}
They can be written in the 8-component form as follows:
\begin{mathletters}
\begin{eqnarray}
\left [i \Gamma^\mu \partial_\mu - m\right ] \Psi_{(+)} (x) &=& 0\quad,
\label{psi1}\\
\left [i \Gamma^\mu \partial_\mu + m\right ] \Psi_{(-)} (x) &=& 0\quad,
\label{psi2}
\end{eqnarray}
\end{mathletters}
where
\begin{eqnarray}
\Psi_{(+)} (x) = \pmatrix{\rho^A (x)\cr
\lambda^S (x)\cr}\quad,\quad
\Psi_{(-)} (x) = \pmatrix{\rho^S (x)\cr
\lambda^A (x)\cr}\quad,
\end{eqnarray}
with $\lambda^{S,A} (p^\mu)$,\, $\rho^{S,A} (p^\mu)$ being the
self/anti-self charge conjugate spinors in the momentum representation,
which are defined in ref.~\cite{DVA}.  The interpretation of $\lambda^S$
and $\rho^A$ answering to positive-energy solutions and $\lambda^A$,\,
$\rho^S$, to negative-energy solutions, have been used.\footnote{Let me
remind that the sign of the phase in the field operator is
considered to be invariant if we restrict ourselves by the orthochroneous
proper Poincar\`e group.  This fact has also been used at the stage of
writing the dynamical equations (\ref{11},\ref{12},\ref{13},\ref{14}).}
After writing those papers we got knowing about similar problems which
have been studied in the old papers~\cite{OTH1,OTH2,OTH3,OTH4} from
various viewpoints.  A group-theoretical basis for such constructs has
been proposed by Bargmann, Wightman and Wigner~\cite{Wigner}.

Let us consider the question of gauge transformations for this kind
of states. First of all, the possibility of the $\gamma^5$ phase
transformations has been noted in~\cite{DVO3}.
The Lagrangian~\cite[Eq.(24)]{DVO3}, which (like in the Dirac
construct) becomes to be equal to zero on the solutions of the dynamical
equations,\footnote{The overline implies the Dirac conjugation.}
\begin{eqnarray}
{\cal L} &=& {i\over 2} \left [
\overline{\lambda^S} \gamma^\mu \partial_\mu \lambda^S -
(\partial^\mu \overline{\lambda^S} ) \gamma^\mu \lambda^S +
\overline{\rho^A} \gamma^\mu \partial_\mu \rho^A -
(\partial^\mu \overline{\rho^A}) \gamma^\mu \rho^A +\right.\nonumber\\
&+&\left.\overline{\lambda^A} \gamma^\mu \partial_\mu \lambda^A -
(\partial^\mu \overline{\lambda^A} ) \gamma^\mu \lambda^A +
\overline{\rho^S} \gamma^\mu \partial_\mu \rho^S -
(\partial^\mu \overline{\rho^S}) \gamma^\mu \rho^S \right ] -
\nonumber\\
&-& m \left [ \overline{\lambda^S} \rho^A + \overline{\rho^A} \lambda^S
-\overline{\lambda^A} \rho^S -\overline{\rho^S} \lambda^A \right ]\,
\end{eqnarray}
is invariant with respect to the phase transformations:
\begin{mathletters}
\begin{eqnarray}
\lambda^\prime (x)
\rightarrow (\cos \alpha -i\gamma^5 \sin\alpha) \lambda
(x)\quad,\label{g10}\\
\overline \lambda^{\,\prime} (x) \rightarrow
\overline \lambda (x) (\cos \alpha - i\gamma^5
\sin\alpha)\quad,\label{g20}\\
\rho^\prime (x) \rightarrow  (\cos \alpha +
i\gamma^5 \sin\alpha) \rho (x) \quad,\label{g30}\\
\overline \rho^{\,\prime} (x) \rightarrow  \overline \rho (x)
(\cos \alpha + i\gamma^5 \sin\alpha)\quad.\label{g40}
\end{eqnarray}
\end{mathletters}
Obviously, the 4-spinors $\lambda^{S,A} (p^\mu)$ and $\rho^{S,A}
(p^\mu)$ remain in the space of self/anti-self charge conjugate
states.\footnote{Usual phase transformations like that applied to
the Dirac field will destroy self/anti-self charge conjugacy.
The origin lies in the fact that the charge conjugation operator
is {\it not} a linear operator and it includes the operation of complex
conjugation.} In terms of the field functions $\Psi_{(\pm)} (x)$ the
transformation formulas recast as follows ($\L^5 = \mbox{diag}
(\gamma^5\quad -\gamma^5)$) \begin{mathletters} \begin{eqnarray}
\Psi^{\,\prime}_{(\pm)} (x) \rightarrow \left ( \cos \alpha + i \L^5
\sin\alpha \right ) \Psi_{(\pm)} (x)\quad,\label{g1}\\
\overline\Psi_{(\pm)}^{\,\prime} (x) \rightarrow \overline \Psi_{(\pm)}
(x) \left ( \cos \alpha - i \L^5 \sin\alpha \right )\quad.\label{g2}
\end{eqnarray} \end{mathletters} It is well known that the Dirac theory
for charged spin-1/2 particles does {\it not admit} conventional chiral
transformations. In the meantime, as mentioned by A.  Das and M.
Hott~\cite{Das} ``an interacting fermion theory at high temperature
develops a temperature dependent fermion mass where mass grows with
temperature; \ldots it would appear that a massless, chiral invariant
theory would have its chiral symmetry broken by the temperature dependent
mass~\cite{temper}; \ldots [on the other hand,] one conventionally
believes that the dynamically broken chiral symmetry in QCD is restored
beyond a critical temperature." Furthermore, they investigated this
``apparent conflict" and proposed $m$-deformed {\it non-local} chiral
transformations. Nevertheless, they indicated at the importance of further
study of chiral transformations and their relevance to the modern physics.
Thus, these matters appear to be of use not only from a viewpoint of
constructing the fundamental theory for neutral particles, but
regarding the constructs which {\it do admit} the chiral invariance may
also be useful for understanding the processes in QCD and other modern
gauge models.

So, let us proceed further  with  the local gradient
transformations (gauge transformations) in the Majorana-Ahluwalia
construct. When we are interested in them one must introduce the
compensating field of the vector potential
\begin{mathletters} \begin{eqnarray} && \partial_\mu
\rightarrow \nabla_\mu = \partial_\mu - ig \L^5 A_\mu\quad,\\ &&
A_\mu^\prime (x) \rightarrow A_\mu (x) + {1\over g} \,\partial_\mu \alpha
\quad.  \end{eqnarray} \end{mathletters}
Therefore, equations describing interactions of the $\lambda^{S}$ and
$\rho^{A}$  with 4-vector potential are the following
\begin{mathletters} \begin{eqnarray}
i\gamma^\mu \partial_\mu \lambda^S (x)
-g\gamma^\mu \gamma^5 A_\mu \lambda^S (x)- m\rho^A (x) &=&
0\quad,\label{i1}\\ i\gamma^\mu \partial_\mu \rho^A (x) +g\gamma^\mu
\gamma^5 A_\mu \rho^A (x) - m\lambda^S (x) &=& 0\quad.\label{i2}
\end{eqnarray} \end{mathletters}
The second-order equations follow
immediately form the set (\ref{i1},\ref{i2})
\begin{mathletters}
\begin{eqnarray} \left \{\left (i \widehat \partial + g \widehat A
\gamma^5 \right ) \left (i \widehat \partial - g \widehat A \gamma^5
\right ) -m^2 \right \} \lambda^S (x) &=& 0\quad,\\ \left \{\left (i
\widehat \partial - g \widehat A \gamma^5 \right ) \left (i \widehat
\partial + g \widehat A \gamma^5 \right ) -m^2 \right \} \rho^A (x) &=&
0\quad;
\end{eqnarray} \end{mathletters}
with the notation being used: $\widehat a \equiv
\gamma^\mu a_\mu = \gamma^0 a^0 - ({\bbox \gamma} \cdot {\bf a})$.
After algebraic transformations in a spirit of~\cite{Itzyk,Ryder}
one obtains
\begin{mathletters} \begin{eqnarray}
\left \{ \Pi_\mu^+ \Pi^{\mu\,+}
-m^2 - {g\over 2} \gamma^5 \Sigma^{\mu\nu} F_{\mu\nu} \right \} \lambda^S
(x) &=& 0\quad,\label{ii1}\\
\left \{ \Pi_\mu^- \Pi^{\mu\,-} -m^2 +
{g\over 2} \gamma^5 \Sigma^{\mu\nu} F_{\mu\nu} \right \} \rho^A (x) &=&
0\quad,\label{ii2} \end{eqnarray}
\end{mathletters} where  the ``covariant derivative" operators
acting in the $(1/2,0)\oplus (0,1/2)$ representation are defined
\begin{equation}
\Pi_\mu^\pm = {1\over i} \partial_\mu \pm g\gamma^5 A_\mu\quad,
\end{equation}
and
\begin{equation}
\Sigma^{\mu\nu} = {i\over 2} \left [ \gamma^\mu\, ,\, \gamma^\nu \right
]_- \quad.
\end{equation}
The case of $\lambda^A$ and $\rho^S$ is very similar and we
shall give below the final result only:  \begin{mathletters}
\begin{eqnarray} \left \{ \Pi_\mu^+ \Pi^{\mu\,+} -m^2 - {g\over 2} \gamma^5
\Sigma^{\mu\nu} F_{\mu\nu} \right \} \lambda^A (x) &=&
0\quad,\label{ii3}\\ \left \{ \Pi_\mu^- \Pi^{\mu\,-} -m^2 + {g\over
2} \gamma^5 \Sigma^{\mu\nu} F_{\mu\nu} \right \} \rho^S (x) &=&
0\quad.\label{ii4} \end{eqnarray} \end{mathletters} Thus,
the equations for the particles
described by the field operator (Eq.  (46) in~[1c])
\begin{equation}
\nu^{^{DL}} (x) \equiv \int \frac{d^3 {\bf p}}{(2\pi)^3} \, {1\over
2p_0} \sum_\eta \left [ \lambda^S_\eta (p^\mu) a_\eta (p^\mu) \exp
(-ip\cdot x) +\lambda_\eta^A (p^\mu) b_\eta^\dagger (p^\mu) \exp (+ip\cdot
x) \right ]\quad,
\end{equation}
which interact with the 4-vector potential,
have the same form for positive- and
negative-energy parts.  The same is true in the case of the use of the
field operator composed from $\rho^A$ and $\rho^S$. One can see the
difference with the Dirac case; namely, the presence of $\gamma^5$
matrix in the ``Pauli term" and in the lengthening derivatives. Next, we
are able to decouple the set (\ref{ii1},\ref{ii2},\ref{ii3},\ref{ii4}) for
the up- and down- components of the bispinors in the coordinate
representation.  For instance, the up- and the down- parts of the
$\nu^{^{DL}} (x) =\mbox{column} (\chi \quad \phi )$ interact with the
vector potential in the following manner:
\begin{eqnarray} \cases{\left
[\pi_\mu^- \pi^{\mu\,-} -m^2 -{g\over 2} \sigma^{\mu\nu}
F_{\mu\nu} \right ] \chi (x)=0\quad, &\cr
\left [\pi_\mu^+ \pi^{\mu\,+} -m^2
+{g\over 2} \widetilde\sigma^{\mu\nu} F_{\mu\nu} \right ] \phi (x)
=0\quad, &\cr}\label{iii}
\end{eqnarray}
where already one has $\pi_\mu^\pm =
i\partial_\mu \pm gA_\mu$, \, $\sigma^{0i} = -\widetilde\sigma^{0i} =
i\sigma^i$, $\sigma^{ij} = \widetilde\sigma^{ij} = \epsilon_{ijk}
\sigma^k$.  Of course, introducing the operator composed of the $\rho$
states one can write corresponding equations for its up- and down-
components  and, hence, restore the Feynman-Gell-Mann
equation~\cite[Eq.(3)]{Feynman} and its charge conjugate
($g = -e$; $A^\mu$ and $F^{\mu\nu}$ are assumed to be real fields).  In
fact, this way would lead us to the consideration which is identical to
the recent papers~\cite{Robson}. It was based on the linearization
procedure for 2-spinors, which is similar to that used by Feshbach and
Villars~\cite{Fesh} in order to deduce the Hamiltonian form of the
Klein-Gordon equation.  Some insights in the interaction issues with the
4-vector potential in the eight-component equation have been made there:
for instance, while explicit form of the wave functions slightly differ
from the Dirac case, the hydrogen atom spectrum is the same to that in the
usual Dirac theory~\cite[p.66,74-75]{Itzyk}.  Next, like in the
paper~\cite{Giesen} the equations of~\cite{Robson} presume a
non-CP-violating\footnote{This is possible due to the Wigner ``doubling"
of the components of the wave function.} electric dipole moment of the
corresponding states.

We are also interested in finding other forms of
gauge interactions for spinors of the $(1/2,0)\oplus (0,1/2)$
representation.  Indeed, one can propose other kinds of phase
transformations and, hence, other  compensating fields for
fermion functions composed from $\lambda^{S,A} (p^\mu)$ and $\rho^{S,A}
(p^\mu)$ spinors.  First of all, one may wishes to introduce the $2\times
2$ matrix $\Xi$, which is defined
\begin{equation} \Xi = \pmatrix{e^{i\phi} & 0\cr 0 &
e^{-i\phi}\cr}\quad,
\end{equation} where $\phi$ is the azimuthal angle
associated with ${\bf p} \rightarrow {\bf 0}$. This matrix has been used
in the generalized Ryder-Burgard relation connecting 2-spinor and its
complex conjugate in the zero-momentum frame (Eq. (26,27) of~[1c]).
Using the relation $\Xi  \Lambda_{_{R,L}} (\overcirc{p}^\mu \leftarrow
p^\mu) \Xi^{-1} = \Lambda_{_{R,L}}^\ast
(\overcirc{p}^\mu \leftarrow p^\mu)$  it is easy to check that under
the phase transformations
\begin{mathletters} \begin{eqnarray}
\lambda_S^\prime (p^\mu) &=& \pmatrix{\Xi & 0\cr 0 & \Xi\cr} \lambda_S
(p^\mu) \equiv \lambda_A^\ast (p^\mu)\quad,\\
\lambda_S^{\prime\prime} (p^\mu)
&=& \pmatrix{0&i\Xi\cr i\Xi & 0\cr} \lambda_S (p^\mu) \equiv i\gamma^0
\lambda_A^\ast (p^\mu)\quad,\\
\lambda_S^{^{\prime\prime\prime}} (p^\mu) &=&
\pmatrix{0&\Xi\cr -\Xi & 0\cr} \lambda_S (p^\mu)\equiv \gamma^0
\lambda_S^\ast (p^\mu)\quad,\\
\lambda_S^{^{IV}}
(p^\mu) &=& \pmatrix{i\Xi & 0\cr 0 & -i\Xi\cr} \lambda_S (p^\mu) \equiv
-i\lambda_S^\ast (p^\mu)
\end{eqnarray}
\end{mathletters}
bispinors remain in the self charge conjugate space. Analogous relations
for $\lambda_A$:
\begin{mathletters}
\begin{eqnarray}
\lambda_A^\prime (p^\mu) &=& \pmatrix{\Xi & 0\cr 0 & \Xi\cr} \lambda_A
(p^\mu) \equiv \lambda_S^\ast (p^\mu)\quad,\\
\lambda_A^{\prime\prime} (p^\mu)
&=& \pmatrix{0&i\Xi\cr i\Xi & 0\cr} \lambda_A (p^\mu) \equiv i\gamma^0
\lambda_S^\ast (p^\mu)\quad,\\
\lambda_A^{^{\prime\prime\prime}} (p^\mu) &=&
\pmatrix{0&\Xi\cr -\Xi & 0\cr} \lambda_A (p^\mu)\equiv \gamma^0
\lambda_A^\ast (p^\mu)\quad,\\
\lambda_A^{^{IV}}
(p^\mu) &=& \pmatrix{i\Xi & 0\cr 0 & -i\Xi\cr} \lambda_A (p^\mu) \equiv
-i\lambda_A^\ast (p^\mu)
\end{eqnarray}
\end{mathletters}
ensure that the latter retain their property to be in the anti-self
charge conjugate space under this kind of transformations.\footnote{One
should still note that  in the meaning presented here, the $\gamma^5$
transformations ($\lambda^S (p^\mu) \leftrightarrow \pm i\lambda^A
(p^\mu)$, see above) are also transformations with a unitary matrix and
also can be regarded as phase transformations of left- (right-) spinors
with respect to right- (left-) spinors.} Thus, the Majorana-like field
operator ($b^\dagger \equiv a^\dagger$) admits additional phase (and, in
general, normalization) transformations, namely
\begin{equation}
\nu^{^{ML\,\prime}} (x) = [c_0 + i({\bbox \tau} \cdot {\bf c})]
\nu^{^{ML\,\dagger}} (x)\quad,\label{pht}
\end{equation} where $c_\alpha$
are arbitrary parameters in the superpositions of the self/anti-self charge
conjugate states; the $\bbox{\tau}$ matrices are defined over the
field of $2\times 2$ matrices;\footnote{It is implied that
$\gamma^0 \equiv {\bbox \tau}_1 \otimes \openone$,\, $\gamma^i \equiv
-i{\bbox\tau}_2 \otimes \sigma^i$ in the Weyl representation of the
$\gamma$ matrices.} and the Hermitian conjugation operation is implied
over the field of the $q-$ numbers, i.~e.  it acts on the $c$-numbers as
the complex conjugation. If we want to keep the normalization of the wave
functions one can make parametrization of the $c_\alpha$ factors in
(\ref{pht}) as follows: $c_0 = \cos \phi$ and ${\bf c} = {\bf n} \sin
\phi$ leaving only three parameters independent.  This induces
speculations that the $SU(2)\times U(1)$ theory can be constructed on the
basis of the Weyl 2-spinors.  This is not surprising, because
these groups are the subgroups of the extended Poincar\'e group. But, of
course, in order to ensure this purpose one should consider the question
of invariance of some Lagrangian, which involves $\lambda$ and $\rho$
fields (e.g., Eq. (24) in ref.~\cite{DVO3}), with respect to these
transformations.

Several forms of field  operators were defined in
ref.~\cite{DVA}; one may be interested in the one composed of
$\lambda^{S,A}$ spinors, and in the second one, of $\rho^{S,A}$ spinors.
Due to the identities (see Eqs.  (6a,6b) in ref.~\cite{DVO4}
and~\cite{DVA})
\begin{mathletters} \begin{eqnarray} \rho^S_\uparrow
(p^\mu) &=& -i\lambda^A_\downarrow (p^\mu)\quad,\quad \rho^S_\downarrow
(p^\mu)= +i\lambda^A_\uparrow (p^\mu)\quad,\quad\\ \rho^A_\uparrow (p^\mu)
&=& +i\lambda^S_\downarrow (p^\mu) \quad,\quad \rho^A_\downarrow (p^\mu) =
- i\lambda^S_\uparrow (p^\mu)\quad,\quad, \end{eqnarray} \end{mathletters}
which permits one to keep the parity invariance of the theory, we can
express the $\rho$ operator in the form:
\begin{equation}
\rho (x^\mu)
\equiv \int {d^3 {\bf p} \over (2\pi)^3 } { 1\over 2p_0 } \sum_\eta \left
[ \rho^A_\eta (p^\mu) c_\eta (p^\mu) \exp (-ip\cdot x) + \rho^S_\eta
(p^\mu) d^\dagger_\eta (p^\mu) \exp (+i p\cdot x) \right ] = \gamma^0
\nu^{^{DL}} (x^{\mu\,\prime})\, .
\end{equation}
The following notation
is used:  $x^{\mu^\prime} \equiv (x^0, - {\bf x})$, and $c_\eta (p^\mu)
\equiv a_\eta (p^{\mu^\prime})$,\, $d_\eta (p^\mu) \equiv  b_\eta
(p^{\mu^\prime})$.
Therefore, the Lagrangian density (24) of ref.~\cite{DVO3} can recast
\begin{eqnarray}
{\cal L} &=& {i\over 4} \left [ \overline{\nu (x)} \gamma^\mu \partial_\mu
\nu (x)  - \partial_\mu \overline{\nu (x)} \gamma^\mu \nu (x) +
\overline{\nu^c (x)} \gamma^{\mu} \partial_\mu \nu^c (x) -
\partial_\mu \overline{\nu^c (x)} \gamma^{\mu} \nu^c
(x) + \right ] +\nonumber\\ &+& {m
\over 2} \left [ \overline{\nu (x)} \gamma^5 \gamma^0 \nu^c (x^\prime)
+ \overline{\nu^c (x)} \gamma^5 \gamma^0 \nu (x^\prime) \right ]
+ ({\bf x} \rightarrow -{\bf x}) \quad. \label{Lagrangian}
\end{eqnarray}
The terms with  ${\bf x} \rightarrow -{\bf x}$
would contribute to the action ${\cal S}$ in the same way as the first
part of Eq. (\ref{Lagrangian}). Therefore, they
can be disregarded.  In subsequent works we shall present the properties
of the this Lagrangian density implied as a
$c$- number with respect to the transformations
(\ref{pht}).  At the moment one can speculate for the $q$- number
theory that, since Eq.  (\ref{pht}) presents indeed themselves the
transformations of the inversion group, which transform to the $q$-
Hermitian conjugate field, and the Lagrangian density usually presents
itself a scalar with respect to both $q$- and $c$- numbers, any $CPT$
invariant theory, which accounts for two types of fields, would be
intrinsically a theory admitting non-Abelian phase transformations of the
components of the field operator.

In the connection with the present work one would wish to pay attention to
the old papers~\cite{Tokuoka}.  Surprisingly, remarkable insights in the
general structure of the $(1/2,0)\oplus (0,1/2)$ representation space and
corresponding interactions have been made as long as thirty years ago but,
unfortunately, this paper also (like several other important works
which I cite here and in my previous papers) remained unnoticeable.
Finally, one can find probable relations between this construct and
that which was used recently, by Moshinsky and Smirnov~\cite{Mosh}.
The latter is based on the concept of the {\it sign spin} of Wigner (the
generators ${\bbox \tau}$, which answer for this concept, were applied for
in many works until the present).

The main conclusion of the paper is: the constructs are permitted,
which are based only on the 4-spinors of the Lorentz group and
which admit the non-Abelian type of phase transformations and, hence,
may admit interactions  of corresponding fields with non-Abelian fields.
If so, this assertment can serve as a basis for explanation of physical
nature of isospin and weak  isospin. Another non-Abelian construct has
been found recently by Profs.  M.  Evans and J.-P.  Vigier in another
representation of the group, and from very different
positions~\cite{Evans}.\footnote{It is still required rigorous development
of the Evans-Vigier ${\bf B}^{(3)}$ construct because it contains several
errors and notational misunderstandings.  Nevertheless, I note some
interesting ideas there and think that one can work rigorously in such a
framework.}

\acknowledgements
I greatly appreciate discussions and information from Profs. D. V.
Ahluwalia, M. W. Evans, Y. S. Kim, A. F. Pashkov, Yu. F. Smirnov  and G.
Ziino.  I am grateful to Zacatecas University, M\'exico, for a professor
position.  This work has been partly supported by the Mexican Sistema
Nacional de Investigadores, the Programa de Apoyo a la Carrera Docente and
by the CONACyT, M\'exico under the research project 0270P-E.

\end{document}